# Polish device for FOCCoS/PFS slit system

Antonio Cesar de Oliveira[1], Ligia Souza de Oliveira[2], Marcio Vital de Arruda[1], Lucas Souza Marrara[2], Leandro Henrique dos Santos[1], Décio Ferreira[1], Jesulino Bispo dos Santos[1], Rodrigo de Paiva Vilaça, Josimar Aparecido Rosa, Laerte Sodré Junior[3], Claudia Mendes de Oliveira[3.]

1- MCT/LNA –Laboratório Nacional de Astrofísica, Itajubá - MG – Brazil
2- OIO-Oliveira Instrumentação Óptica LTDA – SP - Brazil
3- IAG/USP – Instituto de Astronomia, Geofísica e Ciências Atmosféricas/ Universidade de São Paulo - SP – Brazil

## ABSTRACT

The Fiber Optical Cable and Connector System, "FOCCoS", for the Prime Focus Spectrograph, "PFS", is responsible for transporting light from the Subaru Telescope focal plane to a set of four spectrographs. Each spectrograph will be fed by a convex curved slit with 616 optical fibers organized in a linear arrangement. The slit frontal surface is covered with a special dark composite, made with refractory oxide, which is able to sustain its properties with minimum quantities of abrasives during the polishing process; this stability is obtained This stability is obtained by the detachment of the refractory oxide nanoparticles, which then gently reinforce gently the polishing process and increase its efficiency. The surface roughness measured in several samples after high performance polishing was about 0.01 microns. Furthermore, the time for obtaining a polished surface with this quality is about 10 times less than the time required for polishing a brass, glass or ceramic surface of the same size. In this paper, we describe the procedure developed for high quality polishing of this type of slit. The cylindrical polishing described here, uses cylindrical concave metal bases on which glass paper is based. The polishing process consists to use grid sequences of 30μm, 12μm, 9μm, 5μm, 3μm, 1μm and, finally, a colloidal silica on a chemical cloth. To obtain the maximum throughput, the surface of the fibers should be polished in such a way that they are optically flat and free from scratches. The optical fibers are inspected with a microscope at all stages of the polishing process to ensure high quality. The efficiency of the process may be improved by using a cylindrical concave composite base as a substrate suitable for diamond liquid solutions. Despite this process being completely by hand, the final result shows a very high quality.

## 1. INTRODUCTION

We are interested in optimizing the polishing process of the slit device for FOCCoS, "Fiber Optical Cable and Connector System", subsystem of PFS, "Prime Focus Spectrograph", that will be installed in the Subaru telescope.[01] This slit follows a new concept applied in the construction of optical fibers slit device, which ensures the right direction of the fibers by using masks with micro holes. This kind of mask is made by a technique called electroforming, which is able to produce a nickel plate with a liner array of holes. The estimated precision is around 1μm in the diameter and 1μm in the positions of the holes. The design uses two flexible masks, which we call Front Mask and Rear Mask, separated by a gap that defines the thickness of the slit. The pitch and the diameter of the holes define the linear geometry of the slit; the curvature of each mask defines the angular geometry of the slit. The internal structure combines INVAR 36 and a high performance green composite, to ensure good thermal stability inside a range of temperature between 03 °C and 20 °C.

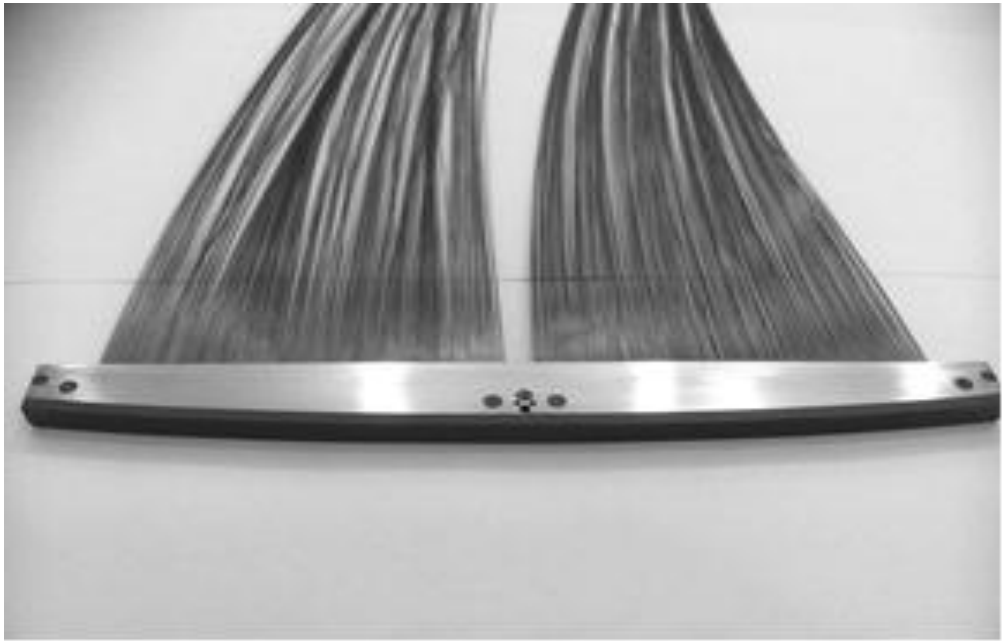

**Figure 1:** Prototype of slit device for FOCCoS/ PFS, following the correct geometry and angular distribution of fibers required by the project.

## 1.1 Slit design

Inside the slit box there are around 616 optical fibers arranged in a linear pattern and laterally positioned, following a standard pitch defined by small holes in both masks of Nickel, as shown in Fig. 2a and 2b. Epoxy or silicone glue may fill the internal space of the slit box keeping the positioning for optical fibers inside the metallic box. The external green composite layer and the optical fibers tips are polished at the same time. The dark green composite is obtained from a mixture of EPO-TEK 301-2, ceramics and other materials in nano-particles form.

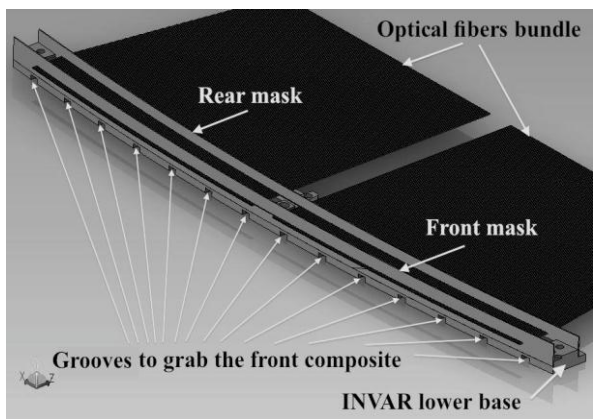

**Figure 2a:** Optical fibers arrangement inside the slit during the first step of building.

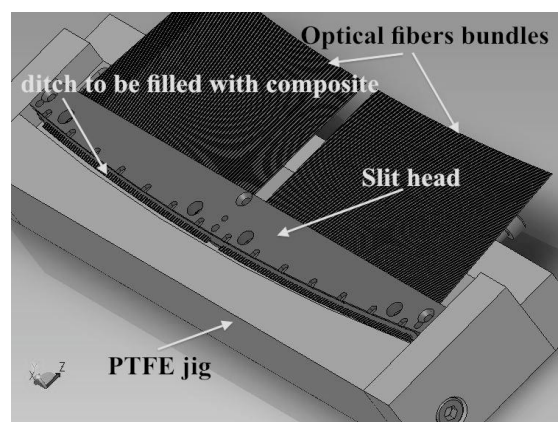

**Figure 2b:** The jig support for composite filling

To avoid bubbles and points of stress, this mixture must be prepared in a separate receptacle inside a vacuum centrifuge machine. The resulting material is more resistant and harder than EPO-TEK 301-2 and, it is well suited for the fabrication of optical fiber arrays. An important secondary characteristic is the ease with which it can be polished. This feature is a result of the presence of micro particles, which keep the polished surface very homogeneous during the final polishing procedure.[02]

## 1.2 Slit specifications

Some specific devices are needed to assure that positioning of each component is done in correct and precise way, as defined according to schematic view showed in Fig. 3. The main requirements are:

1- Gap between the masks: 6 mm
2- Pitch of two consecutive holes in the front mask: 213.930-µm ± 5 µm
3- Pitch of two consecutive holes in the rear mask: 211.840-µm ± 5 µm
4- Radius of curvature of the front mask: 661.5 mm
5- Radius of curvature of the rear mask: 655.5 mm
6- Thickness of the dark composite front cover: 0.5 mm

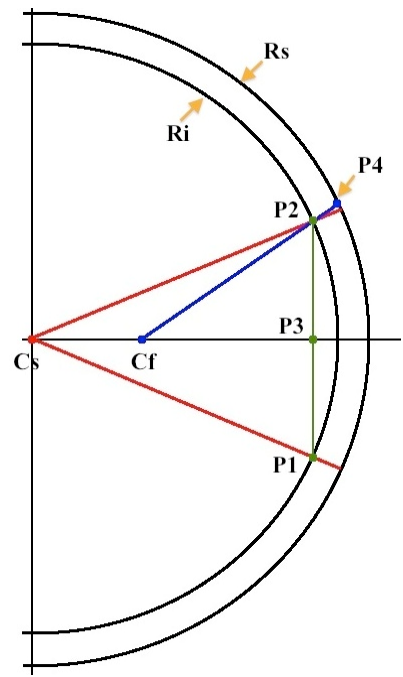

Cs = (Xs,Ys)    **Curvature center for both masks**

Cf = (Xf,Yf)    **Convergence point for the fibers**

P1 = (X1,Y1)    **Intersection point, fiber 1 with mask 1**

P2 = (X2,Y2)    **Intersection point, fiber 2 with mask 1**

P3 = (X3,Y3)    **Midpoint of the vector P1P2**

P4 = (X4,Y4)    **Intersection point, fiber 1 with mask 2**

Ri  = Curvature radius of rear

Rs = Curvature radius of front

**Measuring P1 and P2, it is possible do the calculation of P4 such that Cf, P2 and P4 constituting a straight line.**

**Figure 3:** Mathematical schematic of the slit device for FOCCoS/ PFS

## 2. POLISHING PROCESS

The green composite is particularly useful for optical fibre holders because it contains refractory oxide nanoparticles. These nanoparticles are released during polishing, and gently increase the polishing efficiency. The surface roughness measured in some samples, after high performance polishing, was about 0.01 microns. Furthermore, the time for obtaining a polished surface to this quality is about 10 times less than the time required to polish a surface of brass of the same size.

## 2.1 Polishing mechanism

The Fig. 4a shows the polishing mechanism to be used in curved slits. This kind of mechanism provides a figure 8 polishing pattern on on a substrate with a cylindrical hollow. For the prototype tests we have used a substrate of hard aluminum with the concavity covered with glass paper. This concave base has the same radius of curvature of the slit, 662 mm ± 6 mm. To hold the slit during the polishing movements we developed a special support jig shown in Fig 4b.

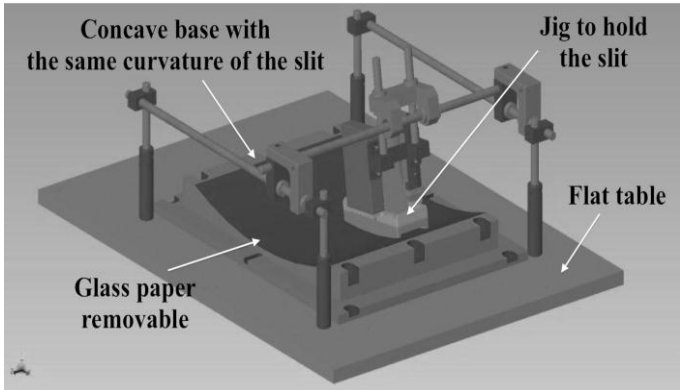

**Figure 4a:** Polishing system developed to polish the slit prototypes, using a grid of glass papers.

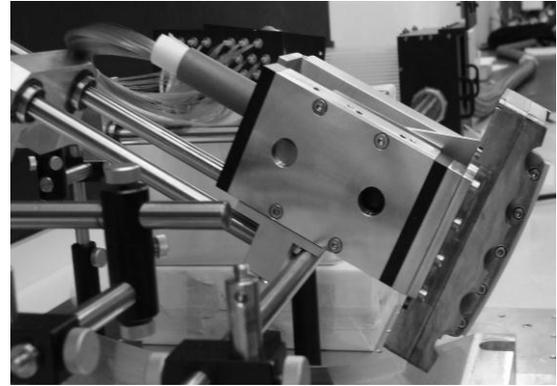

**Figure 4b:** Jig to hold the slit prototypes during the polishing movements.

This jig has a system of springs exerting contact pressure of the slit against the polishing substrate. The backpressure on the front surface of the slit is guaranteed by the weight of the jig. A ball bearing mechanism ensures a smooth displacement in the direction of displacement of the springs. The contact extremity of the jig is constructed in brass following a toothed design so that soap and water can flow between the teeth and spread on the sheet of the glass paper. The set slit and jig is able to move in the shape of a number 8 by through the 3 axes of the polishing mechanism. Each one of the 3 axes consists of a system of linear bearings. The process needs to be done manually but happens very smoothly and can be very easily led with water

## 2.2 Polishing steps

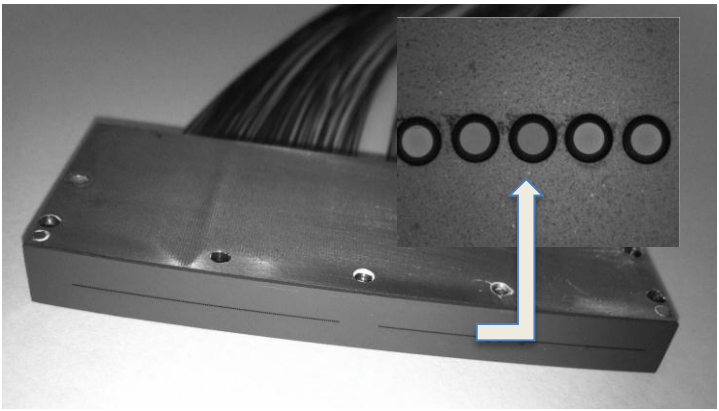

**Figure 5:** Photo of the slit prototype after polishing. In the upper right, a microscopic photo of the fibers shows a closeup of the polished optical fibers.

The polishing process for the green composite covering the front surface of the slit takes several steps. The first step, which removes the excess green composite, use 30um then 12um emery paper in aluminum oxide film. Next, grinding of the cylinder is accomplished using a sequence 9μm, 5μm, 3μm, and 1μm. Final polishing is performed using silica colloidal on a chemistry cloth on the metal curved base. After each step of polishing, it is necessary using a microscope to monitoring the optical fibers surface to insure the high quality of polishing. The Fig. 5 shows a microscopic photo of a region of the fiber array. We have polished small prototypes using a glass paper with good results.

A second polishing process, using only diamond slurry, is under development. With this process, it is possible to start lapping with 3-micron diamond slurry on a copper plate. A second lapping with 1-micron diamond slurry on a tin-lead plate is needed before the final polishing. The final polishing is made with diamond slurry submicron 0.3um. Both plates will be constructed by ENGIS HYPREZ® Lapping & Polishing Systems.

Furthermore, the polishing jig extremity will be constructed with ceramic blocks following properly the curvature of the track polishing.

## 3. POLISHING MATERIALS FOR FOCCoS/PFS SLIT

For this work we expect to use an optical lab with a clean room controlled environment at 23 degrees Celsius, relative humidity 60%, and system of deionized water, countertops, polishing machines, inverted microscopes, magnifiers microscopes and pure compressed air refrigerator dry. These facilities make the LNA Polish Lab an ideal place to fabricate fiber arrays.

### 3.1 Polishing materials
Polish film papers (glass paper) by 3M – 261 lapping film, 30um, 9um.
Diamond in slurries from Hyprez (Engis) Water-soluble, 3um, 0.5um.
Special tooling to support the film papers aluminum concave base.
Hyprez colloidal silica.
Hyprez cleaners' water
Deionized water.
Alcohol Isopropyl

### 3.2 Concave track polishing
On this polish we intend to use 3 types of support concave bases, Fig. 6, with 400X600mm HY TX10A in aluminum, copper and tin lead. The estimated curvature is around 662 mm. The format of them is the negative format that we need to have on the curved slit. Working with Engis HYPREZ® Lapping & Polishing Systems we create a unique polish system to help on this polish.

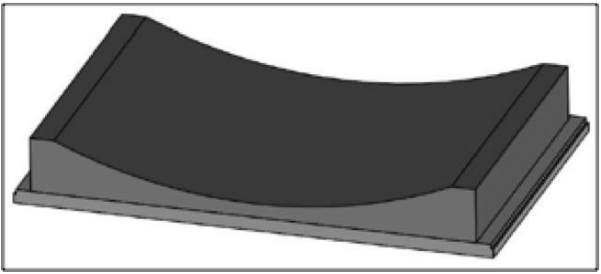

**Figure 6:** The figure shows the format of the support concave bases. 400 x 600 mm and radius of curvature of 662 mm

### 3.3 Tools - Ceramic blocks
The extremity of the polishing jig is a tool, shown in Fig. 7, to help deposit the diamond on the concave copper and tin bases to distribute the abrasive uniformly. This tooling is made of hard ceramic and the pressure weight of it makes pressing at the surface of the plates to distribute or deposing the diamond sully in the copper or tin lead polishing track.

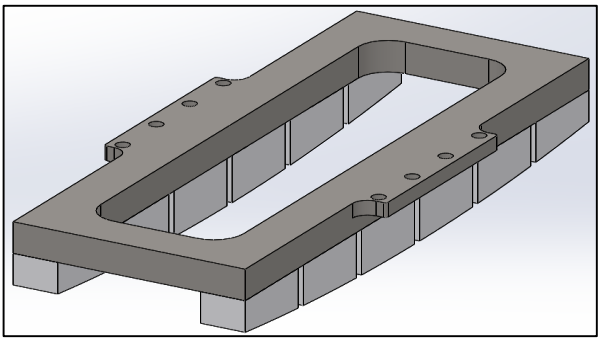

**Figure 7:** Ceramic conditioning tool 95 mm x 200 mm, convex backed with a radius of curvature adjusted with the polishing track.

### 3.4 Reconditioning of the concave base. Diamond tooling

The recondition of the concave plate is a part of the polish system that opens the porous in the concave base to receive the diamond slurry. However, the greatest importance of this tool is to eventually be used to correct small deviations of the radius of curvature of the concave polishing track caused by irregular drainage of material from the polishing track by the abrasives.

### 3.5 Polishing jig

After the Slit is molded and cured, it must be assembled onto the polishing jig shown in Fig. 8. This jib ensures optimal slit positioning in X, Y, Z and perpendicularity for the complete alignment. These preliminary tasks have direct impact on final geometry of the slit, and ensure uniform thickness of the polished composite.

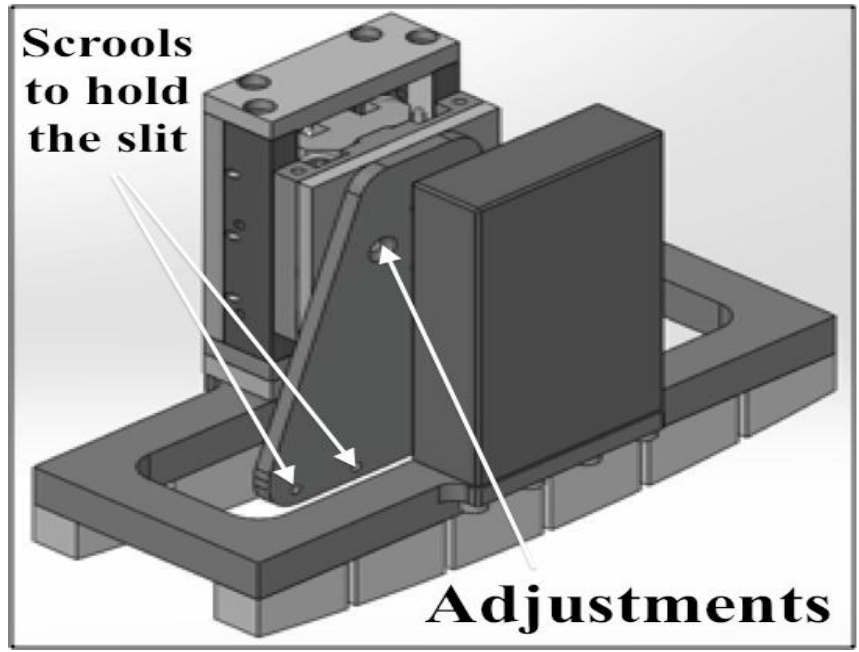

**Figure 8:** Polishing jig with ceramic block at the extremity to match with the curvature of the polishing concave track base.

## 4. METROLOGY AND QUALITY CONTROL

In the current phase of this project, we are developing metrology to evaluate the quality of polishing as well as quantifying parameters after completion of polishing. While the image analysis with optical microscopy should be sufficient to evaluate the quality of the polish, we are investigating the best options for measuring the radius of curvature without damaging the polished surface. This research points to two possible ways forward; the first is the acquisition of non-invasive optical metrology equipment and the second is the acquisition of stock analysis software that can be incorporated into our toolmakers microscope. In any event, a mechanical goniometer and contact profilometer have shown very good results for measuring the radii of curvature within the require precision. However special care needs should be taken to avoid contamination of the environment including polishing abrasives to be used. In general, companies that provide these materials are very careful to ensure products with high purity, but the wrong handling can be a source of contamination.

# 5. SUMMARY AND CONCLUSIONS

We describe in this article, the cylindrical polishing process that will be used to polish the slit devices of the instrument FOCCoS / PFS to be installed on the Subaru telescope. This process was successfully tested in slit prototypes and appropriate implementations point to future improvements. Although it is a manual process, we expect to achieve high quality and precision using professional tools and abrasives custom high purity. Procedures still in the testing phase must be validated by a non-invasive metrology, yet to be incorporated in this work. However, initial results from conventional mechanical metrology show very good results.

# 6. ACKNOWLEDGMENTS

We gratefully acknowledge support from: Fundação de Amparo a Pesquisa do Estado de São Paulo (FAPESP), Brasil. Laboratório Nacional de Astrofísica, (LNA) e Ministério da Ciência Tecnologia e Inovação, (MCTI), Brasil. We would like also to gratefully Daniel J. Reiley from Caltech-California, for help us with the correction of this paper. Finally we are very grateful to INCT-A (Instituto Nacional de Ciência e Tecnologia - Astrofisica) to fund our participation in SPIE.